\documentclass[sigconf]{acmart}

\AtBeginDocument{%
  }

\setcopyright{acmlicensed}
\copyrightyear{2025}
\acmYear{2025}
\setcopyright{rightsretained}
\acmConference[CHI EA '25]{Extended Abstracts of the CHI Conference on Human Factors in Computing Systems}{April 26-May 1, 2025}{Yokohama, Japan}
\acmBooktitle{Extended Abstracts of the CHI Conference on Human Factors in Computing Systems (CHI EA '25), April 26-May 1, 2025, Yokohama, Japan}\acmDOI{10.1145/3706599.3719730}
\acmISBN{979-8-4007-1395-8/2025/04}

\usepackage{cleveref}
\usepackage{xcolor}

\definecolor{myblue}{rgb}{0.216,0.494,0.965}

\newcommand{\new}[1]{\textcolor{black}{#1}}

\newcommand*\highmlhighb[0]{\textcolor{black}{\small{\texttt{HM-HD}}}}
\newcommand*\lowmllowb[0]{\textcolor{black}{\small{\texttt{LM-LD}}}}
\newcommand*\lowmlhighb[0]{\textcolor{black}{\small{\texttt{LM-HD}}}}
\newcommand*\highmllowb[0]{\textcolor{black}{\small{\texttt{HM-LD}}}}

\newcommand*\lowml[0]{\textcolor{black}{\small{\texttt{LM}}}}
\newcommand*\highml[0]{\textcolor{black}{\small{\texttt{HM}}}}
\newcommand*\lowb[0]{\textcolor{black}{\small{\texttt{LD}}}}
\newcommand*\highb[0]{\textcolor{black}{\small{\texttt{HD}}}}

\begin{document}

\title[Interactivity x Explainability]{Interactivity x Explainability: Toward Understanding How Interactivity Can Improve Computer Vision Explanations
}

\author{Indu Panigrahi}
\orcid{0009-0002-2084-4756}
\affiliation{%
  \institution{Princeton University}
  \city{Princeton}
  \state{New Jersey}
  \country{USA}
}
\email{indup@princeton.edu}

\author{Sunnie S. Y. Kim}
\authornote{Authors contributed equally.}
\orcid{0000-0002-8901-7233}
\affiliation{%
  \institution{Princeton University}
  \city{Princeton}
  \state{New Jersey}
  \country{USA}
}
\email{sunniesuhyoung@princeton.edu}

\author{Amna Liaqat}
\authornotemark[1]
\orcid{0000-0002-5170-1945}
\affiliation{%
  \institution{Princeton University}
  \city{Princeton}
  \state{New Jersey}
  \country{USA}
}
\email{al0910@princeton.edu}

\author{Rohan Jinturkar}
\orcid{0009-0001-4818-5636}
\affiliation{%
  \institution{Princeton University}
  \city{Princeton}
  \state{New Jersey}
  \country{USA}
}
\email{rohanj@alumni.princeton.edu}

\author{Olga Russakovsky}
\orcid{0000-0001-5272-3241}
\affiliation{%
  \institution{Princeton University}
  \city{Princeton}
  \state{New Jersey}
  \country{USA}
}
\email{olgarus@princeton.edu}

\author{Ruth Fong}
\orcid{0000-0001-8831-6402}
\affiliation{%
  \institution{Princeton University}
  \city{Princeton}
  \state{New Jersey}
  \country{USA}
}
\email{ruthfong@princeton.edu}

\author{Parastoo Abtahi}
\orcid{0009-0000-2145-3445}
\affiliation{%
  \institution{Princeton University}
  \city{Princeton}
  \state{New Jersey}
  \country{USA}
}
\email{parastoo@princeton.edu}

\renewcommand{\shortauthors}{Panigrahi et al.}

\begin{abstract}
    Explanations for computer vision models are important tools for interpreting how the underlying models work. However, they are often presented in static formats, which pose challenges for users, including information overload, a gap between semantic and pixel-level information, and limited opportunities for exploration. We investigate interactivity as a mechanism for tackling these issues in three common explanation types: heatmap-based, concept-based, and prototype-based explanations. We conducted a study (N=24), using a bird identification task, involving participants with diverse technical and domain expertise. We found that while interactivity enhances user control, facilitates rapid convergence to relevant information, and allows users to expand their understanding of the model and explanation, it also introduces new challenges. To address these, we provide design recommendations for interactive computer vision explanations, including carefully selected default views, independent input controls, and constrained output spaces. 
\end{abstract}




\maketitle

\section{Introduction}
AI explanations help users understand and diagnose complex AI models, such as modern computer vision (CV) models processing thousands of pixels \cite{adadi2018xaisurvey, barredo2020xaisurvey}.
While the explainable AI (XAI) community has developed many explanation methods, they often present information in a static form which can be challenging for users to interpret \cite{kriglstein2014gameplay, xia2018heatmap, blignaut2010heatmap, kim2023hmhta, ramaswamy2023overlooked, kim2016criticisms}. Moreover, a single static explanation cannot meet the diverse needs of users with varying expertise \cite{liao2020xaidesign}. \looseness=-1

Most XAI methods produce static CV explanations \cite{gradcam, lime, petsiuk2018rise, fong2017meaningful, brendel2019bagnet, simonyan2013deep, shitole2021sag, zeiler2014visualizing, Zhou2016CAM, koh2020concept, ramaswamy2022elude, shukla2023cavli, zhou2018ibd, chen2019protopnet}. However, recent studies have identified several challenges with static explanations \cite{kim2023hmhta, kriglstein2014gameplay, xia2018heatmap, blignaut2010heatmap, ramaswamy2023overlooked, kim2016criticisms} which we categorize into three groups: 

\begin{enumerate}
    \item \textbf{Information overload}: Users feel that there is too much information to process at once \cite{kriglstein2014gameplay, xia2018heatmap, blignaut2010heatmap, kim2023hmhta, ramaswamy2023overlooked, kim2016criticisms}.
    
    \item \textbf{Semantic-pixel gap}: Users find it difficult to connect image pixels to objects, attributes, \new{and abstract concepts} \cite{ramaswamy2023overlooked, kim2023hmhta, blignaut2010heatmap}.
    
    \item \textbf{Limited exploration}: Users \new{lack the necessary tools} to probe the explanation to deepen their understanding \cite{kim2023hmhta}.
\end{enumerate}



Prior work in Information Visualization has shown that interacting with data helps with interpretation \cite{keim2002data, kim2017infovisinteractivity, yi2007interaction, hohman2019gamut}, and consequently, several works have called for interactivity in XAI \new{applications} \cite{Kulesza2015InteractiveML, sundar2012interactiveeffective, sundar2010interactivitycustom, Nguyen2024CVPRW, slack2023talktomodel, hohman2019gamut}. 
However, interactivity has mostly been explored at the dataset level \cite{yi2007interaction, boggust2022embedding, hohman2019gamut, fogarty2008cueflik, bauerle2023symphony, wang2023wizmap, dudley2018mlinterfacesurvey}, for visualizing model weights \cite{wang2021cnnexplainer, chang2016appgrouper, kapoor2010manimatrix, hohman2020summit, dudley2018mlinterfacesurvey}, and for non-visual data \cite{liu2021slider, slack2023talktomodel, Bhattacharya2023Directive, Bhattacharya2024exmos}. In this work, we explore interactivity \textit{at the explanation level} for images.
Specifically, we investigate two research questions: \textbf{RQ1}: How do end-users \textbf{leverage interactivity} to help understand information conveyed by CV explanations? \textbf{RQ2}: How do end-users \textbf{perceive interactive CV explanations}? \looseness=-1

To answer these questions, we conducted a within-subjects study using explanations for a bird identification model \cite{cubdataset}. We selected bird identification over common object or scene classification use cases \cite{places365, imagenet} to examine the effects of domain knowledge. We focused on three widely used explanation types (heatmap-based, concept-based, and prototype-based explanations) and investigated three interactive mechanisms: \textbf{Filtering} to control the amount of information, \textbf{Overlays} to connect pixel-level landmarks to semantic labels, and \textbf{Counterfactuals} to allow users to edit images and observe a change in the explanation. We recruited 24 participants with varying levels of machine learning (ML) and birding expertise for an $\sim90$-minute study. Each participant completed a set of tasks, alongside a questionnaire and interview.\looseness=-1

We found that interactivity addresses some limitations of static CV explanations by providing users with tools to bridge the semantic-pixel gap, rapidly pinpoint the information they seek, explore the CV model beyond the immediate task, and gain clarity around the presentation of the static explanation. However, some interactive mechanisms were at times overwhelming for users and introduced new challenges, which informed the design recommendations we present in this paper. Our exploration and study findings contribute to our understanding of interactive mechanisms for XAI and bring us closer to the design of effective interactive CV explanations. \looseness-1

\section{Related Work}
\label{sec:related}
Given the increasing ubiquity of AI \cite{singhal2023medpalm, ma2024medsam, huang2023plip, janai2020autonomous, michielssen2024baseball}, there exists a growing need for end-users to understand a model's behavior so they can appropriately trust and use the model.
To address this need, many XAI methods have been developed over the past decade across various subfields of AI, \new{including} CV and robotics~\cite{Lundberg2017shap, danilevsky2020nlpxai, lime, milani2024xrlsurvey, qian2021nlpxaiinterface}, and across different domains, \new{such as }healthcare and climate science~\cite{DeGrave2023auditing, antoniadi2021clinical, Lotsch2022Biomedicine, Singh2020Imaging, Smith2021ClinicalAI, Atakishiyev2021autonomous, Meteier2019AutomatedDriving, Machlev2022Energy, Mamalakis2022Climate}.
Historically, much work has focused on \textit{developing methods} to explain a model's outputs (i.e., \new{generating an answer for} \textit{``why did the model \new{make this prediction?}''}).
Recently, research has emerged at the intersection of XAI and Human-Computer Interaction (HCI) on \textit{evaluating how useful} explanation methods are for users with varying machine learning (ML) expertise \cite{nauta2023xaievalsurvey, jinli2022medxaihci, nazar2021medxaihci, kim2022hive}, which we build on for CV explanations. 
In this section, we highlight literature most related to our focus on interactive CV explanations.




\subsection{Computer Vision Explanations}
\label{sec:cv_explanations_problems}
In this work, we focus on \emph{attribution-based} explanations, which compute a measure of importance by identifying the parts of an input that are critical for a model's decision \cite{goyal2019counterfacvis, miller2021contrastive} and have been the most developed in XAI research thus far.
We discuss three types of attribution-based explanations.
\textit{Heatmap-based explanations} typically generate a temperature heatmap that indicates the importance of each pixel or image region for the model's prediction~\cite{gradcam, lime, petsiuk2018rise, fong2017meaningful, brendel2019bagnet, simonyan2013deep, shitole2021sag, zeiler2014visualizing, Zhou2016CAM}.
\textit{Concept-based explanations} explain a model's output by assigning a numerical importance scores to semantic concepts (e.g., +2.7 for ``wings'')~\cite{koh2020concept, ramaswamy2022elude, shukla2023cavli, zhou2018ibd}.
\textit{Prototype-based explanations} learn a set of important image regions (i.e., prototypes) from training images \cite{chen2019protopnet, rymarczyk2022proto, kenny2023towards, nauta2021prototree, donnelly2022deformable, Nguyen2022team}.\looseness=-1


Computer vision explanations have primarily been static.
However, several studies have highlighted three main issues with static explanations.
The first is \textit{information overload.} Prior work has found that the amount of information in explanations can be overwhelming. With heatmaps, users often want to view the ``extremes'' of a heatmap and find intermediate colors distracting~\cite{kriglstein2014gameplay, xia2018heatmap, blignaut2010heatmap}. For concepts, users report feeling overwhelmed by the number of items \cite{kim2023hmhta, ramaswamy2023overlooked}, often the number of annotated objects in a dataset (e.g., $1197$ in the Broden dataset~\cite{bau2017netdissect}). Similarly, prototypes can require many regions to explain a model well enough \cite{kim2016criticisms}.
The second is a \textit{semantic-pixel gap.} Prior work suggests users often find it difficult to connect image regions to the represented object and vice versa. For concepts, users have difficulty identifying the concepts in the images \cite{ramaswamy2023overlooked}. With heatmaps and prototypes, users struggle to discern what objects are in the highlighted regions \cite{kim2023hmhta, blignaut2010heatmap}.
The third is a \textit{limited means for user-driven exploration.} 
Although attribution-based explanations show \textit{what} regions are important, they do not explain \textit{why} they are important \cite{miller2021contrastive, wachter2018contrastive, bekri2020expapproach}. Indeed, Kim et al. \cite{kim2023hmhta} found that users wanted to know why certain regions were deemed important by exploring the underlying causal relationships.\looseness=-1
\subsection{Interactivity and Explainable AI}
While the problems presented could be partially mitigated by adjusting the amount of information in the static explanation on a case-by-case basis, interactivity has been identified as an effective \cite{sundar2012interactiveeffective, kim2017infovisinteractivity, sundar2010interactivitycustom, srinivasan2019interactive} and customizable \cite{sundar2010interactivitycustom, srinivasan2019interactive} way for users to interact with and understand data \cite{boggust2022embedding, yi2007interaction}.
Much of the existing work in interactivity for XAI has largely implemented interfaces that allow users to explore images at the dataset level by filtering through data points based on model predictions or their calculated similarity \cite{hohman2019gamut, boggust2022embedding, fogarty2008cueflik, bauerle2023symphony, wang2023wizmap, dudley2018mlinterfacesurvey}.
Other work provides tools to examine the inner workings of models (e.g., model weights and learned features) \cite{feng2023uxforml, guo2022interactiveeditrules, wang2021cnnexplainer, chang2016appgrouper, kapoor2010manimatrix, hohman2020summit, dudley2018mlinterfacesurvey}. 
There is limited research on interactivity at the explanation level for a single image. Some related work includes interactivity for textual or tabular data \cite{liu2021slider, slack2023talktomodel, Bhattacharya2023Directive, Bhattacharya2024exmos}.
For example, Liu et al. \cite{liu2021slider} explore interactive explanations that rely on direct manipulation, such as sliders and drop-down menus.
However, they focus on comparing how human-AI teams with interactive explanations perform compared to AI-only agents when identifying out-of-distribution points.
Within computer vision, Nguyen et al. \cite{Nguyen2024CVPRW} develop an interface for human-AI teams where users provide feedback on explanations.
In contrast, our work seeks to investigate how users leverage and perceive interactive CV explanations and to provide design recommendations.\looseness=-1






\section{Explorative User Study}

\new{In this section, we describe the details of our IRB-approved study.}
\subsection{Study Method}

\subsubsection{\new{Terminology}}
\begin{figure*}[ht]
    \centering
    \includegraphics[width=0.95\textwidth]{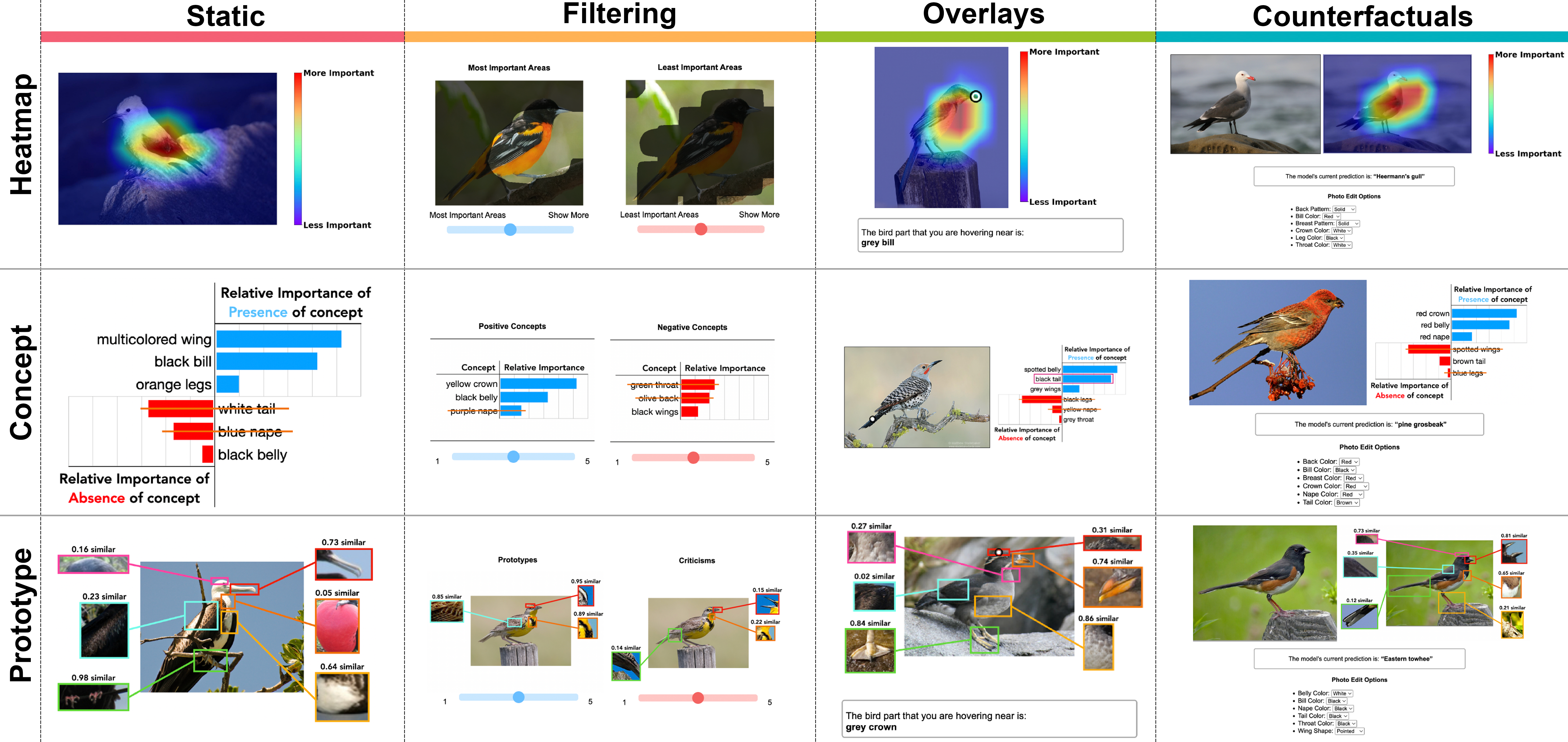}
    \caption{12 explanation mock-ups for 3 explanation types (rows) and 4 presentation types (columns). All bird images were from the Caltech-UCSD Birds-200-2011 dataset~\cite{cubdataset}.}
    \Description{Table with 4 columns are labeled "Static", "Filtering", "Overlays", and "Counterfactuals" and 3 rows labeled "Heatmap", "Concept", and "Prototype". Each cell in the table contains a picture.}
    \label{fig:all_12_mockups}
    \vspace{-0.1cm}
\end{figure*}

We tested three \textit{explanation types}: heatmap-\new{based}, concept-\new{based}, and prototype-based (i.e., the rows in \cref{fig:all_12_mockups}).
We created mock-up explanations, allowing us to carefully control the \textit{presentation} of explanations across participants.
For each explanation type, we created four \textit{presentation types}: Static, Filtering, Overlays, and Counterfactuals (i.e., the columns in \cref{fig:all_12_mockups}).
The latter three were interactive mechanisms inspired by the problems outlined in~\Cref{sec:cv_explanations_problems}.
An \textit{explanation} is a pairing of one explanation type and one presentation type (i.e., one cell in \cref{fig:all_12_mockups}).\looseness-1

\subsubsection{Setup}
For the \textbf{Static} baseline (\cref{fig:all_12_mockups}, 1st col), we based our mock-ups on prior works~\cite{gradcam,chen2019protopnet,kim2023hmhta}. 
We then adapted them so that they shared some uniform characteristics across presentation types.
First, we wanted all explanations to have a visual component, so we used bar graphs instead of numerical scores in the concept version (heatmap and prototype versions already had visual components), similar to TCAV~\cite{Kim2018tcav}.
This design decision was also in line with prior work, which reported that users dislike or feel overwhelmed by the numerical coefficients \cite{kim2023hmhta}.
Second, we wanted explanations to describe an individual image (i.e., local explanation), not an entire output class (i.e., global explanation) which in our case was a bird species.
Concept-based explanations are typically global, as concepts may not appear in every image.
Thus, we made local concept-based explanations by crossing out concepts not present via strikethrough~\cite{kim2023hmhta}.\looseness=-1

\textbf{Filtering} (\cref{fig:all_12_mockups}, 2nd col) is a type of interactivity that allows the user to conditionally remove information~\cite{figueiras2015interactivitytypes, keim2002data, yi2007interaction, shneiderman1996datareviewinteractivity}.
We utilized sliders that hide or show components of the explanation, conditioned on their assigned importance~\cite{liu2021slider}.
In the concept and prototype versions, the sliders remove or add concepts and prototypes.
In the heatmap version, we replaced the color gradient with a binary mask that grows or shrinks as the slider is moved. 
We chose this design because layering the mask on top of the color gradient makes the underlying image difficult to see~\cite{kim2023hmhta}.\looseness-1

\textbf{Overlays} (\cref{fig:all_12_mockups}, 3rd col), described by Shneiderman as ``details-on-demand''~\cite{shneiderman1996datareviewinteractivity}, are extra layers of information~\cite{kong2012overlays} that appear only when the user indicates that they want to view them \cite{vermette2017exoverlays}.
We incorporate a hover-over that supplies the pixel locations or the semantic labels of bird parts in the explanation, depending on which is missing.
When users hover over a heatmap or prototype version, a dot appears on the closest bird part and a dialog box below the explanation shows a textual label for that part.
For the concept version, when users hover over a concept in the bar chart that is visible in the image (i.e., not crossed out), a dot representing the concept's location appears on the image.\looseness=-1

\textbf{Counterfactuals} (\cref{fig:all_12_mockups}, 4th col) explain model predictions by allowing the user to explore how changes to model inputs affect model outputs~\cite{miller2023decisionxai}.
Typically, static counterfactual explanations show \new{an edited image with the smallest modification needed to} change the model's output prediction from one class to another (e.g., how does an image need to change so that a model predicts ``Cardinal'' instead of ``Blue jay''?)~\cite{goyal2019counterfacvis,vandenhende2022counterfactual}.
We utilize counterfactuals as a form of interactivity by allowing users to edit an image and see how the explanation and prediction subsequently change.
Users are provided with six drop-down menus that we refer to as ``edit options''.
Each menu corresponds to a bird part and provides two options for the \new{attribute} of that bird part: one is the original and the other is a randomly selected alternative.
For example, if the bird has a red wing and the selected alternative is a blue wing, clicking the menu for ``wing'' will display the options ``red'' and ``blue''.
If a user selects the alternative \new{attribute}, the image updates to show the edit, and the explanation updates when the edited image changes it.
Additionally, we include a dialog box that displays the model's prediction for the current image; this also updates if the edited image changes the model's prediction.\looseness-1

For each explanation, we chose a different image to mitigate learning effects. We used a series of preprocessing steps and consulted with two birding experts to select 12 images of similar difficulty in terms of \new{bird} identification and image quality.
We performed these checks to ensure \new{some level of consistency across images in terms of} lighting, camera angle, or other factors. Additionally, two CV experts (two authors) checked that the explanations were of similar difficulty.

\subsubsection{Participants}
We designed a within-subjects study and recruited 24 participants with varying levels of ML and domain expertise (see Appendix A).
We advertised our study through various platforms, including several institutions' email lists, birding labs, online conservation and birding forums, X, and Mastodon. Due to our snowball sampling process, all participants held an academic or research affiliation.
Studies were conducted over Zoom video calls and lasted 90--100 minutes. 
All participants received a $\$25$ gift card.\looseness=-1

\subsubsection{Procedure}
Each participant tried all 12 explanations in \Cref{fig:all_12_mockups}.
We formed a balanced Latin square across the 3 explanation types and randomized the presentation type ordering so that we had 4 participants per ordering, one from each expertise category.
For each explanation, the participants explored and interacted with the explanation mock-up in a web application and then provided written answers \new{to the survey questions} (see Appendix B, images of the interface are included in our supplementary material.).
Specifically, participants were first asked to read the instructions and then view and interact with the explanation.
The instructions described the explanation type and how to use the interactive mechanism when it was present.
The participant could choose when to proceed to the survey, though we enforced a time limit of roughly two minutes.
The survey questions asked participants to identify the most and least important parts of a bird based on the explanation; these tasks required participants to actively engage with each explanation.
All task questions included a reference to a bird part guide that we created to assist with birding terminology. 
At the end of the study, participants rated their level of agreement with a set of statements and ranked the presentation types in terms of their general preference and preference in the context of learning about a new bird species, allowing for ties.\looseness=-1

\subsubsection{Analysis}
We collected qualitative data from the study recordings and quantitative data from the survey questions.
For our qualitative data, we created a codebook from the audio and video recordings of the studies and performed a Reflexive Thematic Analysis \new{\cite{braun2019reflex, Braun2006qual}} to extract findings and outline recommendations.
Some codes directly came from verbal responses (e.g., ``\textit{User prefers counterfactuals for learning because exploration}''), while other codes described participant interactions (e.g., ``\textit{User interacted with each of the edit options individually}'').
Two authors created an initial codebook based on four studies; one of those authors coded the remaining studies.
All authors iterated on the codebook.
Since one author coded all the studies and refined the codebook in agreement with the other authors, we did not calculate inter-rater reliability.\looseness=-1 



\subsection{Study Results}
    We present the quantitative results from the surveys followed by the themes from our qualitative analysis of the interviews, summarize the takeaways, and offer design recommendations.\looseness=-1

\begin{figure}[b]
    \centering
    \includegraphics[width=1\columnwidth]{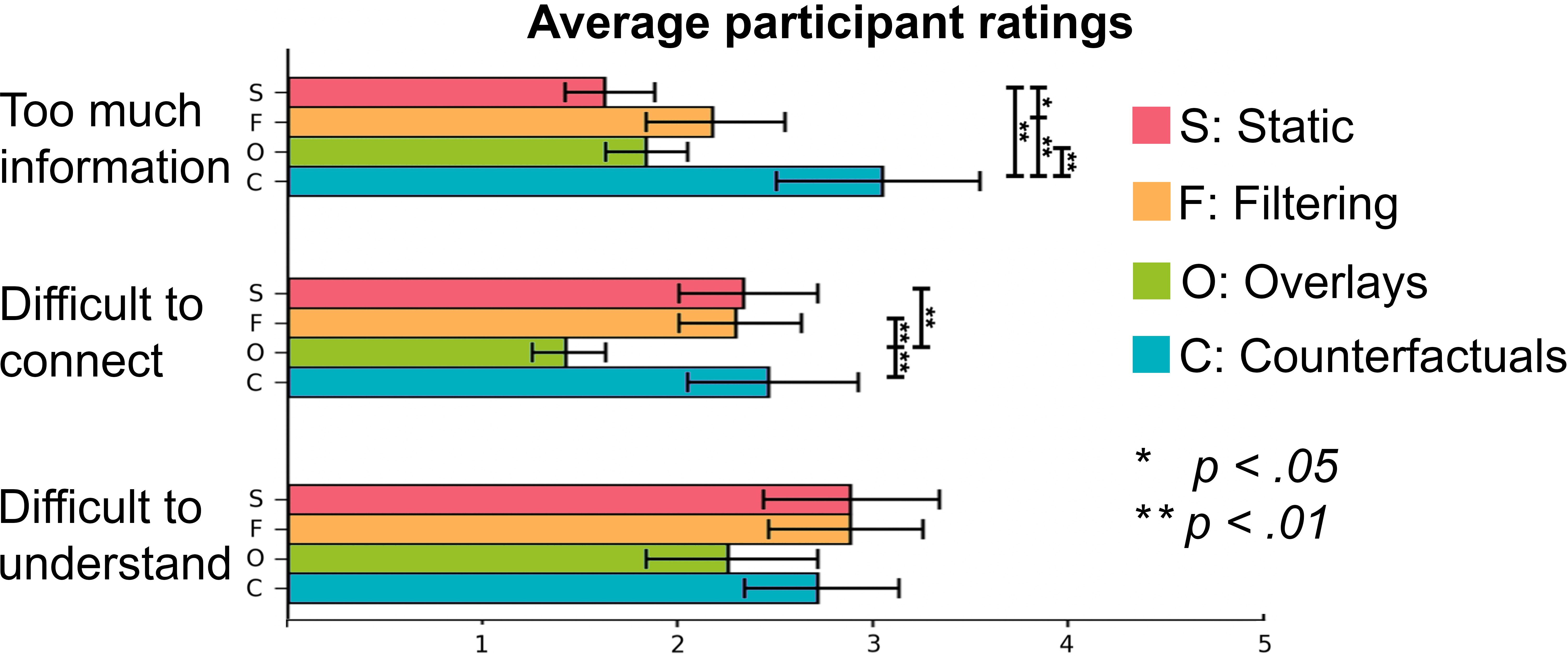}
    \caption{Average participant ratings for survey statements. Lower is better. Error bars are 95\% confidence intervals.}
    \Description{Average Ratings on Problem-related Statements.}
    \label{fig:ratings_results}
\end{figure}

\subsubsection{Questionnaire Results}
After viewing all 12 explanations, participants rated a set of statements (see Appendix B) on a 5-point Likert scale and ranked the four presentation types. 
Lower numbers indicate better ratings (\cref{fig:ratings_results}).
For the analysis, we used a two-sided Wilcoxon signed-rank test with Holm correction; we used Holm to avoid assuming independence.
Participants rated Counterfactuals as having significantly more information compared to Static, Filtering, and Overlays ($p < .01$). Participants rated Filtering as having significantly more information than Static ($p = .015$). 4 participants noted that while Filtering provided more information than Static, the additional information was not overwhelming. For connecting pixel-level information to semantic-level information, participants rated Overlays as significantly less difficult than Static, Filtering, and Counterfactuals ($p < .01$). There were no significant differences in ratings for the model understanding statement. In general, participants preferred Overlays significantly more than Static ($p < .01$) and Counterfactuals ($p = .02$). For learning about bird species, participants preferred Overlays significantly more than Static presentations ($p < .01$). \looseness-1

\begin{figure}[b]
    \centering
    \includegraphics[width=1\columnwidth]{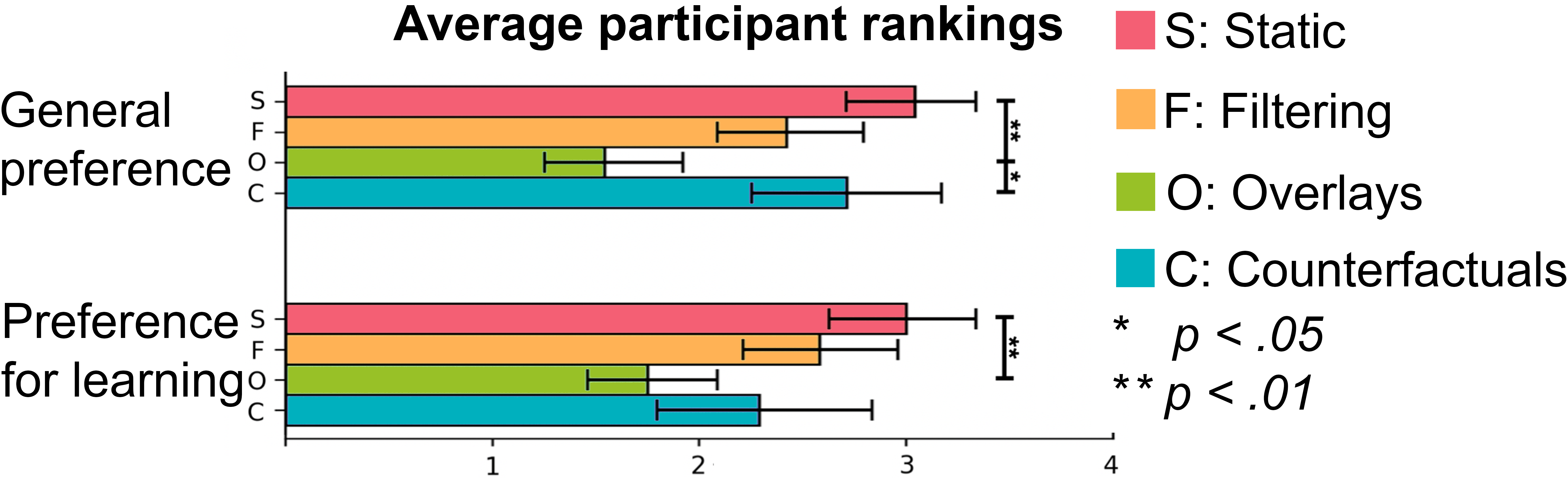}
    \caption{Average participant rankings for general preference and preference for learning bird species. Ties are allowed. Lower is better. Error bars are 95\% confidence intervals.}
    \Description{Average participant rankings for general preference and preference for learning bird species.}
    \label{fig:rankings_results}
\end{figure}


\subsubsection{Qualitative Results}
    Here, we discuss the qualitative themes we developed. Parentheses indicate the number of participants. \looseness-1

\vspace{0.1cm}
\noindent\textbf{\textit{Participants appreciate interactive mechanisms that augment the explanation without changing the underlying explanation.}} 
Participants appreciated Filtering and Overlays as tools for adjusting the amount of detail in the explanation without changing the explanation itself. Specifically, participants mentioned that Filtering made explanations less overwhelming by giving them control over the amount of visible information (12) and this control in turn made explanations easier to digest (4). Participants also used this interactive mechanism to reduce the amount of information (8) or gradually add information (7) during the task. Similarly, participants liked Overlays (18) despite the additional information presented. For example, P2 said: ``\textit{[Overlays have] more information than [Static], but I feel like the way they're presented makes it so that you can take it in easier, and it doesn't feel like more information}''. On the other hand, participants found Counterfactuals overwhelming (12), as image edits altered the underlying explanations.

\vspace{0.1cm}
\noindent\textbf{\textit{Participants find interactive mechanisms that alter the underlying explanation overwhelming.}}
    Participants found that Counterfactuals were overwhelming because there could be simultaneous changes in the explanation when the image was modified. \new{Additionally, they} struggled to untangle the causal relationships because the inputs were interdependent (11). P6 commented that ``\textit{In my mind, it’s hard to reason about the combination of influences}'', and P2 stated for the prototypes version that they ``\textit{think there's too much going on with the photo editing and the [prototypes] popping out}''. In fact, participants often performed one edit at a time, making sure to revert each edit option to the default choice before trying another (61 out of 72 Counterfactuals trials). Moreover, participants had to explore a large number of possible explanations that resulted from the 64 possible edited images (9). P7 remarked: ``\textit{There were so many things to look at---it was confusing}''. Overall, participants found Counterfactuals confusing and hard to interpret (12), particularly those with low ML and domain expertise (5), and needed more time to explore the space of possible inputs and \new{the resulting} outputs (6).\looseness-1 

    
\vspace{0.1cm}
\noindent\textbf{\textit{Although participants find Counterfactuals overwhelming, they utilize them to resolve confusion around static presentations by inducing systematic changes in model predictions and explanations.}}
    Participants leveraged Counterfactuals to clarify aspects of static explanations they found confusing and to confirm their understanding of the explanations. For example, many participants were confused about static concept-based explanations (16), and some used Counterfactuals to edit the image and observe the resulting changes in the model's prediction and explanation to clarify their understanding, which ultimately led to a more accurate interpretation of the static explanation (6). P17 noted: ``\textit{It was helpful to be able to change the color of [a bird part] and see how that affected the bar graph...that sort of helped me understand what the [strikethrough] is for}''. Additionally, the visual changes helped participants identify the bird parts (15). Hohman et al. observed similar behaviors where “\textit{...participants used Counterfactuals often throughout their exploration, both as a direct task and as a sanity check for feature sensitivity}” in their study with ML practitioners \cite{hohman2019gamut}.\looseness-1

\vspace{0.1cm}
\noindent\textbf{\textit{Participants leverage Counterfactuals to explore a range of explanations to better understand the AI model more broadly and beyond the task at hand.}}
    Many participants (20) found Counterfactuals useful for understanding the AI model's reasoning at a high level. Some participants expressed a desire to use Counterfactuals to explore other birds outside of the one image given and to build a macroscopic understanding of the AI model. Specifically, participants appreciated the ability to change the bird species by editing the image (6). This ability to explore allowed participants to develop an understanding of the model more broadly. For example, P15 thought it helped them build rules about how changes in the model’s input affect its output, and P18 noted that this mechanism would allow them to identify the decision points of the model. These uses were beyond the scope of the tasks which only asked for the most and least important bird parts in the given image. This suggests that participants saw Counterfactuals as a source of information to explore and expand their understanding of the model, even if they were not asked to do so.\looseness=-1

\vspace{0.1cm}
\noindent\textbf{\textit{Participants felt that Filtering and Overlays allowed them to quickly focus on information of interest.}}
    Some participants appreciated that Filtering automatically sorted the bird parts by importance (e.g., similarity scores sorted by magnitude in prototype-based explanations) because it obviated the need for manual comparison (6). For instance, P6 liked the prototype version of Filtering because ``\textit{it produces a ranking of the similarity scores that I don't need to think about myself}''. Similarly, participants found Overlays useful for easily and quickly identifying the names of the bird parts, even with the bird part guide provided (8). For example, P21 remarked that Overlays were ``\textit{faster because I could hover over it and it could tell me what the parts were}''. Thus, participants felt that the automatic sorting and labeling offered by Filtering and Overlays respectively enabled them to efficiently hone in on the information that they were seeking. Participants noted that the lack of these features in Static presentations was inconvenient (5). \looseness=-1

\subsubsection{Summary}

    Participants found that interactive features augmenting explanations, such as Filtering and Overlays, helped them understand information without being overwhelming. While Counterfactuals introduced significant visual changes by altering the underlying explanations, participants used them to clarify confusing aspects of static explanations, reaffirm their understanding, and explore the AI model more broadly. With Filtering and Overlays, participants felt they could quickly focus on relevant information, and they appreciated the flexibility that these mechanisms offered when interpreting explanations.\looseness=-1

            

\subsubsection{Design Recommendations}
Participants expressed that interactive mechanisms were at times confusing or inconvenient. Based on these findings, we recommend the following design considerations for future work on interactive CV explanations.

\vspace{0.1cm}
\noindent\textbf{\textit{Avoid interdependent input controls.}} Participants found Counterfactuals confusing because they could not disentangle the effects of multiple simultaneous edits on the prediction or the explanation. A more appropriate design might involve preventing users from performing more than one edit at a time, thereby clarifying the causal relationship between each input and the corresponding \new{changes}. \looseness-1

\vspace{0.1cm}
\noindent\textbf{\textit{Constrain the input and output space.}} With Counterfactuals, many participants felt overwhelmed by the number of possible image edits and the extent of changes to the explanation. We suggest limiting the input and output space to create a manageable range of information, tailored to the application, allowing participants to easily explore all.\looseness-1

\vspace{0.1cm}
\noindent\textbf{\textit{Design an optimal static default view.}} We recommend that users should have the option to choose whether or not to engage with interactive mechanisms. Therefore, the default static presentation should provide enough information for the initial interpretation of the explanation without overwhelming users \cite{ramaswamy2023overlooked}.\looseness-1

    



\section{Limitations}
Our study is a preliminary step toward evaluating interactive CV explanations and thus has several limitations. For example, our study sessions were relatively short, with participants only having two minutes to explore each explanation. Some participants expressed a desire for more time, suggesting the need for future studies to explore deeper engagement with interactive mechanisms for CV explanations. Additionally, longitudinal studies are needed to understand how users interact across repeated sessions and the role of interactive mechanisms that maintain user history.\looseness-1

We chose a simple bird identification task, which was a reasonable starting point, because it was consistent with prior work \cite{kim2023hmhta, kim2023humans, Morrison2024Imperfect} and allowed us to recruit participants with low and high domain expertise. However, it is unclear to what extent our findings can be generalized to more complex, high-risk, and high-impact applications such as medical diagnostics. Moreover, our sample size was relatively small, underscoring the need for larger-scale studies. Lastly, our interactive mechanisms were not directly comparable to the static explanations. For example, the Filtering mechanism for heatmap-based explanations lacked the color gradient present in its static counterpart. Carefully controlled experiments are needed for quantitative analysis of performance, trust, and confidence in tasks using various interactive computer vision explanations.\looseness-1
\section{Future Work}
Future research should explore a broader range of interactive mechanisms for CV explanations beyond the three examined in this work. Additionally, some participants suggested integrating information beyond the image, such as geographical context or general facts, to improve the utility of explanations. Further work is needed to define the boundaries of what information should be included in interactive explanations across different application domains. Collaborating with domain experts and stakeholders through participatory design will be critical for advancing these directions. 

\new{Finally,} prior work has shown that explanations may lead to misplaced trust in AI systems ~\cite{danry2023overreliance,kim2022hive,kim2025fostering,zhang2020effect,Bansal2021CHI}. Interactivity in explanations could amplify these risks, further misleading users and enabling confirmation bias. For example, users might iteratively interact with Counterfactuals until the explanations appear to align with their preconceptions. More research is needed to carefully examine these possibilities and mitigate potential negative outcomes.\looseness=-1
\section{Conclusion}
    Most computer vision explanations are static, limiting user interaction. Drawing from prior XAI research, we identified three key issues with these explanations: information overload, a semantic-pixel gap, and limited means for user-driven exploration. Given the potential of interactivity in XAI, we explored how users engage with and perceive interactive explanations by comparing three mechanisms across three explanation types. We conducted a study with 24 participants of varying ML and domain expertise. Our findings showed that interactivity helps address the challenges of static explanations by offering users greater control and quicker access to relevant information. While some mechanisms were overwhelming, they allowed users to expand their understanding of the model and refine their grasp of the static explanation itself. Based on these insights, we offered design recommendations for interactive computer vision explanations, including optimized default views, independent input controls, and constrained interaction spaces.\looseness=-1

\begin{acks}
Firstly, we thank our participants for taking the time to participate in our study. We also thank Patrick Newcombe and Fengyi Guo for their help in identifying bird species and images of similar difficulty for the study. Lastly, we thank members of Princeton HCI and the Princeton Visual AI Lab, especially Vikram Ramaswamy, for providing feedback and resources throughout this project.


We also acknowledge support from the NSF Graduate Research Fellowship Program (SK), the Princeton SEAS Howard B. Wentz, Jr. Junior Faculty Award (OR), the Princeton SEAS Innovation Fund (RF), and Open Philanthropy (RF, OR). 
\end{acks}

\bibliographystyle{ACM-Reference-Format}
\bibliography{references}

\appendix
\section*{Appendix}
\section{Expertise Criteria and Distribution}
    We recruited 6 participants from each of the following 4 categories: high-ML and high-domain (\highmlhighb), high-ML and low-domain (\highmllowb), low-ML and high-domain (\lowmlhighb), and low-ML and low-domain (\lowmllowb). 
    Participants rated their familiarity with ML and the task domain of birding on a 5-point scale, and were considered to have high or low expertise based on their responses:
    \begin{itemize}
        \item \lowml: \textit{``1: I don't know anything about ML''}, \textit{``2: I have heard about a few ML concepts or applications''}, or \textit{``3: I know the basics of ML and can hold a short conversation about it.''}\looseness-1
        \item \highml: \textit{``4: I have taken a course on ML and/or have experience working with an ML system''} or \textit{``5: I often use and study ML.''}\looseness-1
        \item \lowb: \textit{``1: I don't know anything about birding''}, \textit{``2: I have heard about a few birding concepts''}, or \textit{``3: I know the basics of birding and can hold a short conversation about it.''}\looseness-1
        \item \highb: \textit{``4: I have taken a course on birding and/or have experience in birding''} or \textit{``5: I often conduct bird-watching and study birding.''}\looseness-1
    \end{itemize}

    \noindent Specifically, the distribution of participants was as follows: 
    \begin{itemize}
        \item \lowmlhighb~ included P11, P12, P15, P16, P19, and P21.
        \item \highmlhighb~ included P10, P14, P17, P18, P23, and P24.
        \item \highmllowb~ included P3, P4, P6, P8, P13, and P20.
        \item \lowmllowb~ included P1, P2, P5, P7, P9, and P22.
    \end{itemize}

\section{Survey Questions}
\noindent \textbf{Q1.} Rate your agreement with the following statements on a 5-point scale from 1 (strongly disagree) to 5 (strongly agree): \looseness-1
    \begin{enumerate}
        \item It was difficult to read the explanation.
        \item There was too much information provided.
        \item It was difficult to connect the explanation to parts of the photo.
    \end{enumerate}

\noindent \textbf{Q2.} Based on the labels in the table, rank your preference of the following four groups of explanation designs (1 being most preferred and 4 being least preferred) and verbally explain your thought process. The labels were W, X, Y, and Z with each letter representing Static, Filtering, Overlays, and Counterfactuals respectively.


\noindent \textbf{Q3.} Let's say that you needed to learn about a new bird species. Rank how useful the four groups of explanation designs would be for doing so (1 being most useful and 4 being least useful) and verbally explain your thought process. As with Q2, the labels were W, X, Y, and Z with each letter representing Static, Filtering, Overlays, and Counterfactuals respectively.


\end{document}